\documentclass{aa}
\usepackage{graphicx}
\usepackage{txfonts}
\usepackage{amssymb}
\usepackage{amsmath}
\usepackage{hyperref}
\usepackage{xspace}
\usepackage{xcolor}
\usepackage{rotating}
\usepackage{caption}
\usepackage{booktabs}
\usepackage{tabularx}
\usepackage{orcidlink}

\let\oldorcidlink\orcidlink
\renewcommand\orcidlink[1]{\smash{\raisebox{.3em}{\large{\oldorcidlink{#1}}}}}
\hypersetup{
  colorlinks,
  allcolors=blue
}
\newcommand{\vp}{\texttt{Visplot}\xspace}
\newcommand{\astropy}{\texttt{Astropy}\xspace}
\newcommand{\pyephem}{\texttt{PyEphem}\xspace}
\newcommand{\astroplan}{\texttt{astroplan}\xspace}
\newcommand{\slalib}{\texttt{SLALIB}\xspace}
\newcommand{\staralt}{\texttt{Staralt}\xspace}
\newcommand{\eqreft}[1]{Eq.~\eqref{#1}}
\newcommand{\eqrefp}[1]{Eq.~\ref{#1}}
\newcommand{\figref}[1]{Fig.~\ref{#1}}
\newcommand{\appref}[1]{Appendix~\ref{#1}}
\newcommand{\panelref}[1]{\figref{fig:overview}/#1}
\newcommand{\secref}[1]{Sect.~\ref{#1}}
\newcommand{\secreftwo}[2]{{Sects.~\ref{#1}} and~\ref{#2}}
\newcommand{\tabref}[1]{Table~\ref{#1}}

\begin{document}
\title{\vp: A visibility plot and observation scheduling tool for astronomical observatories}
\author{Emanuel Gafton\inst{1,2}\corrauth{ega@ing.iac.es}\orcidlink{0000-0003-0781-6638}
   \and Illa R. Losada\inst{1}\email{illa.rivero.losada@gmail.com}\orcidlink{0000-0002-0416-7516}}
\institute{Nordic Optical Telescope, Rambla José Ana Fernández Pérez 7, Breña Baja, 38711 La Palma, Spain
      \and Isaac Newton Group of Telescopes, Apartado 321, Santa Cruz de La Palma, 38700 La Palma, Spain}
\date{Received 13 April 2026 / Accepted 7 September 2026}
\abstract
{}
{We present \vp, a free and open-source tool designed for interactive
construction and refinement of nightly telescope observing schedules through a
graphical user interface. The software performs hardware-aware observability
analysis by explicitly incorporating telescope-specific limits alongside
astronomical constraints, in order to determine feasible observation windows for
both sidereal and non-sidereal targets. It also generates observing schedules
that can be rapidly adapted to changing conditions.}
{\vp is a browser-based JavaScript application that can be used without
installation, while remaining fully customizable through self-hosted deployments.
The software computes target altitudes for the specified observatory coordinates
on a discretized time grid spanning the entire observing night. Observability windows 
for individual targets are then derived by evaluating their altitude profiles 
against the intersection of astronomical constraints (airmass, moon distance,
twilight), mechanical telescope boundaries (altitude and hour-angle limits), and
custom temporal restrictions defined in coordinated universal time (UTC) or 
local sidereal time (LST).
The scheduling engine operates within these observability windows, selecting and
ordering observations using a multi-constraint heuristic optimization scheme.
Mandatory targets are scheduled deterministically, while the remaining target
pool is ranked using a weighted scoring function that balances user-defined
priority, target urgency, altitude, and telescope slew overhead.}
{Originally developed to address operational scheduling requirements at the
Nordic Optical Telescope (NOT), \vp has been in continuous use for a decade
and has grown beyond its original development to support
astronomers at multiple observatories worldwide, demonstrating its practical
value in real-world observational workflows. The software supports rapid
schedule revision and sharing of observing plans, making it well suited for
target-of-opportunity (ToO) observations and geographically distributed remote
observing. User feedback indicates that the tool reduces the effort of nightly
planning while ensuring compliance with telescope operating limits.
The interactive visualization of observing constraints also provides a useful
educational resource for teaching observational astronomy and telescope operations.}
{}
\maketitle
\nolinenumbers

\section{Introduction}\label{sec:intro}
Efficient planning of ground-based astronomical observations requires accurate 
assessment of target observability as a function of time, observatory location, and
observing constraints. Factors such as airmass, telescope altitude limits,
twilight, moon brightness, and moon distance impose strict limits on when
targets may be observed. These factors must be evaluated during the preparation of both
observing proposals and nightly observing plans. While some observatories
maintain internal tools for this purpose, such solutions are generally
site-specific, tightly coupled to local infrastructure, or unavailable outside
their host institutions. As a result, observers frequently rely on a
heterogeneous collection of scripts, web tools, and manual procedures to
evaluate target observability and construct observing schedules.

Several general-purpose astronomy software packages such as \astropy
(\citealp{astropy2022}) and \pyephem (\citealp{rhodes2011}), and
lower-level toolkits including \slalib (\citealp{wallace2014}),
\texttt{PAL} (\citealp{PAL2013}), \texttt{SOFA} (\citealp{wallace2009}),
\texttt{ERFA} (\citealp{erfa2024}), \texttt{libnova} (\citealp{girdwood2015})
and \texttt{AA+}\footnote{\url{http://www.naughter.com/aa.html}}
provide accurate ephemeris calculations and coordinate transformations, but
they do not address the practical needs of nightly observing planning in an
integrated manner.
The \astropy-affiliated package \astroplan (\citealp{morris2018})
extends this functionality to include constraint-based scheduling, but 
remains a software library intended for programmatic use and requiring 
custom scripting to support specific observing workflows, rather than 
providing an integrated, interactive planning environment.

A number of tools offer visualization capabilities for assessing target 
observability. Applications such as \staralt\footnote{\url{https://astro.ing.iac.es/staralt}}
and \texttt{JSkyCalc}\footnote{\url{https://github.com/jrthorstensen/JSkyCalc}}
produce static altitude--time charts, while services like
\texttt{airmass.org}\footnote{\url{https://airmass.org}}
allow for interactive plotting.
More specialized tools, such as \texttt{ViSiON} (\citealp{carry2018})
support observability calculations for both sidereal and non-sidereal targets,
and desktop platforms like Subaru Telescope's \texttt{SPOT}\footnote{\url{https://github.com/naojsoft/spot}}
provide rich, interactive environments including finding charts and real-time
monitoring. However, these tools are focused on visualization and
manual planning, and do not provide automated scheduling.
At the other end of the spectrum, commercial software platforms
for integrated observatory management (such as \texttt{iObserve} and its
web-based successor \texttt{arcsecond.io}\footnote{\url{https://www.arcsecond.io}})
support end-to-end workflows including proposal preparation,
communication with the telescope--instrument control system, 
and queue scheduling. While powerful, these systems generally require some form of
institutional integration and subscription-based access, and are often proprietary, restricting access to the source code and
limiting the extent to which
users can independently customize or extend the software. Neither are they aimed at rapid, interactive
observability assessment and schedule exploration without a dedicated
infrastructure.
Since all of these tools are agnostic to telescope-specific hardware constraints
(such as altitude- or instrument-dependent pointing restrictions),
the observer must manually cross-reference the visual plan
with the physical and safety limitations of their specific facility.
While such constraints could, in principle, be integrated into some of these
frameworks by a motivated user, doing so would necessitate substantial source-code
modification and the development of custom software features.

Taken together, existing open-access tools address individual aspects of observability
analysis, visualization, or scheduling, but do not simultaneously provide 
hardware-aware constraint modelling, integrated scheduling, and real-time 
interactive refinement within a single application
requiring no installation.
These aspects are especially relevant for
small and medium-sized facilities, remote observing sites, collaborative
observing campaigns involving multiple observatories, and educational contexts,
where ease of use and immediate visual feedback are particularly important.
To address this gap, we developed \vp, an observability analysis and 
scheduling tool designed for practical, single-night operations. It
generates visibility charts for multiple targets over the course of a night, similar to
\staralt, whose operational use informed our initial design,
while extending this functionality through the inclusion of constraints, integrated scheduling, 
interactive finding charts, and all-sky camera overlays.
The scheduling engine is designed to produce operationally efficient
solutions that account for user-defined priority weighting and a
range of physical constraints, both target-specific (e.g., airmass limits, moon 
distance, and temporal restrictions) and telescope-specific (e.g., altitude limits),
thereby enabling the transition from simple visibility
assessment to complex, constraint-aware observation planning.
The software supports real-time schedule refinement, 
consistent with the broader literature on astronomical scheduling,
where human-in-the-loop tools that support rapid re-planning are essential due
to the dynamic nature of observing conditions (e.g. \citealt{solar2016}).
It also performs rigorous coordinate transformations and ephemeris 
calculations for both sidereal and non-sidereal targets, and supports automatic 
retrieval of target coordinates or ephemerides.
Finally, because many astronomers routinely observe across multiple facilities,
the ability to rely on a familiar planning tool that functions consistently 
across sites represents a significant operational advantage.

This paper describes \vp version 5.5, which was the latest public release at the time of submission.
\secref{sec:software} presents a software overview of \vp, including its
design, internal module structure, observatory integration capabilities, and
deployment model.
\secref{sec:model} describes the conceptual framework underlying \vp, including 
the mathematical foundations of observability analysis and the scheduling
algorithms used to construct observing plans.
\secref{sec:workflow} outlines the operational workflows supported by the
software (covering nightly planning, proposal preparation, and real-time
observations) and includes an illustrative example of scheduling generation 
for a representative observing night.
\secref{sec:adoption} summarizes the adoption and operational impact of \vp
within the astronomical community based on its first decade of continuous 
service.
Finally, \secref{sec:summary} discusses potential future development paths
and provides concluding remarks.

\section{Software overview}\label{sec:software}
\subsection{Software architecture}\label{software:architecture}
\vp is engineered as a client-side web application. Rather than relying on a
traditional client--server architecture where a backend server processes
requests and performs the necessary calculations, \vp leverages modern web
standards (HTML5, CSS3, and JavaScript following the ECMAScript 5 specifications\footnote{ECMAScript 
5, standardized in 2009, provides broad compatibility with older web browsers,
helping to support legacy computers still in use at some observatories.})
to perform all calculations within the user's browser.
Due to this architecture, the server is primarily responsible for delivering
static application files, and the number of concurrent users is therefore
limited by conventional web-serving capacity rather than by CPU-intensive
calculations. On contemporary laptop hardware, calculations and rendering are
completed within a few hundred milliseconds for typical target lists of up to
$\sim$50 objects, allowing changes to be reflected instantaneously from the user's perspective.
For a list of 50 targets, the computational time is approximately 200~ms,
while memory usage averages 5~MB based on Firefox profiling. Both computation time
and memory usage scale approximately linearly with the number of targets. In practice,
computational performance is not the limiting factor; rather, the graphical presentation
becomes the primary constraint, as the visibility plot and the schedule become
visually crowded when dozens of targets are shown simultaneously, well before CPU
or memory resources become limiting.

\begin{figure}[t!]
\begin{center}
\includegraphics[width=.95\columnwidth]{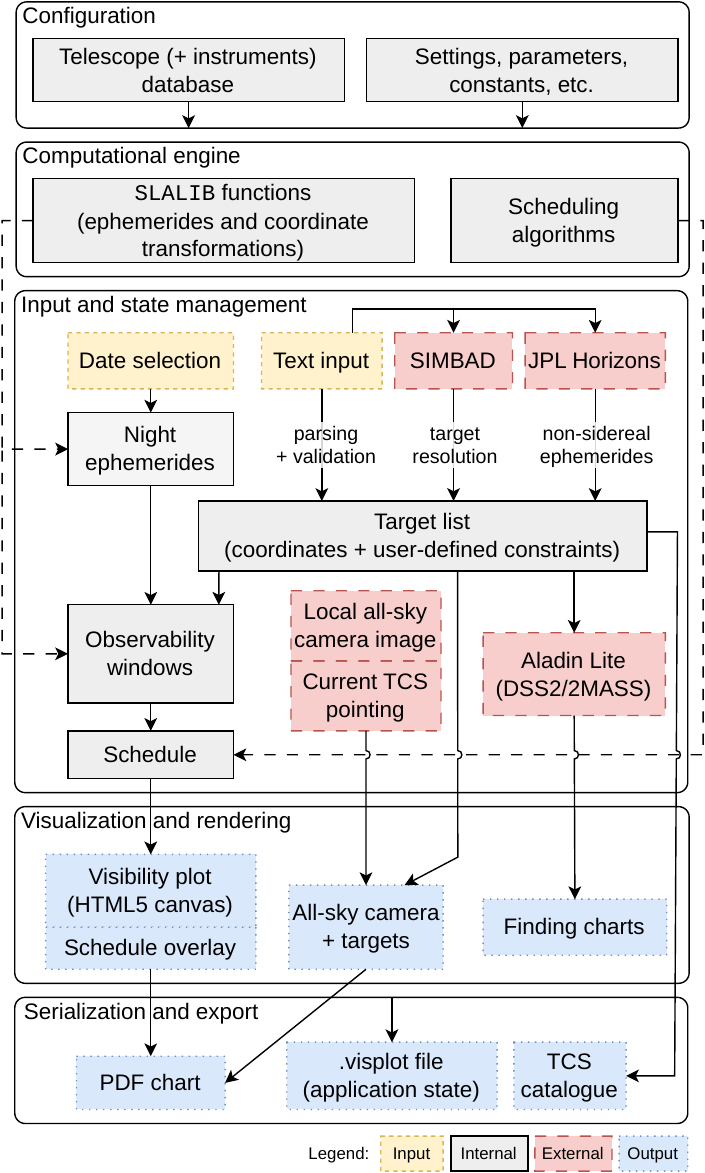}
\end{center}
\caption{Software architecture of \vp, showing the primary modules and their interactions.
Solid arrows indicate primary data flow, while dashed arrows indicate dependencies.
Arrows pointing directly to (for configuration) or from (for serialization) the
system boundary indicate data flow to/from all other parts of the application.}
\label{fig:diagram}
\end{figure}

Once the core assets are loaded, \vp operates mostly offline, allowing users
to modify the target list and schedule even in the event of network connectivity problems. Network requests
are limited to queries to: (i) Centre de données astronomiques de Strasbourg
(CDS), for coordinate lookups in SIMBAD (\citealp{wenger2000}) and for the
creation of finding charts using Aladin Lite v3 (\citealp{baumann2022}),
(ii) the NASA Jet Propulsion Laboratory (JPL) Horizons System (\citealp{giorgini2015}),
for retrieval of non-sidereal target ephemerides, and (iii) telescope-specific
all-sky camera images and telescope control system (TCS) pointing data
(if configured, as discussed in \secref{software:integration} below). Target and schedule
information remain local to the user's device, ensuring strict data privacy.

The codebase is divided into five primary modules that adhere to a strict
separation of concerns, as shown in \figref{fig:diagram}:

\begin{itemize}
\item Configuration. This module contains a meticulously curated database of over 120
optical and near-infrared telescopes. Each entry includes at least the latitude, longitude,
elevation above sea level, and Internet Assigned Numbers Authority (IANA) time zone identifier.
Where available, it also includes telescope and instrument-specific altitude limits,
detector properties (such as field of view and flip, used for displaying finding charts),
and all-sky camera settings (used to retrieve all-sky images and map pixels to alt-azimuth coordinates).
The telescope database is presented through an interactive \texttt{Leaflet}\footnote{\url{https://leafletjs.com}}
map using \texttt{ArcGIS Online} World Imagery tiles with optional cloud and precipitation
overlays from \texttt{OpenWeatherMap}\footnote{\url{https://openweathermap.org}}. 
The map also features a custom day/night
terminator layer that displays the current daytime, twilight, and nighttime regions of the Earth
together with the subsolar point.
The module also centralizes physical constants, visual preferences, parameters, and settings
used by the rest of the code.

\item Computational engine. This is the core logic module. Based on the
observatory configuration and the sanitized target data, it
calculates the night ephemerides and target altitudes, and implements the observability-window construction
and scheduling algorithms described in \secreftwo{model:visibility}{model:scheduling} below.
All coordinate transformations and ephemeris calculations in \vp are performed using standard \slalib
(\citealp{wallace2014}) functions, ported to JavaScript as part of the \vp project itself.
An internal \slalib\ unit-test suite is executed automatically whenever the application is loaded,
verifying the correctness of the JavaScript port while also confirming that the required modules
have been loaded and executed successfully.
IANA time zones (such as \texttt{Atlantic/Canary}) are converted to UTC offsets
(such as WET/UTC+0 or WEST/UTC+1, depending on the time of year) using the
\texttt{Moment.js}\footnote{\url{https://momentjs.com}} library.

\item Input and state management. This component handles data ingestion, including
parsing user-supplied target lists and resolving coordinates and ephemerides through external services.
It maintains the application's global state, including the target list, observing constraints,
observability windows, and generated schedules.
The module coordinates the flow of information between user inputs, the computational engine, and 
the visualization layer using an event-driven approach, in which user
actions update the application state, triggering selective recomputation of dependent
quantities and subsequent visualization updates.

\item Visualization and rendering. To handle the dynamic plotting of dozens of 
visibility curves efficiently, this layer bypasses the standard Document Object Model (DOM) and
uses the low-level HTML5 Canvas API to draw the airmass curves and schedules, together with additional
metadata (axes, altitude constraints, list of targets, and solar/lunar night ephemerides).
Interactivity with this canvas through event-based triggers enables real-time user actions, 
such as the manual drag-and-drop reordering of targets and the instantiation of contextual 
modal popups that display finding charts, instrumental configurations, and target details.
The web page has a responsive design, adapting its layout automatically to phones and tablets
while preserving the same functionality as the desktop version. Although the arrangement of interface
elements is adjusted to suit smaller screens, there is no noticeable impact on performance,
making the application usable across a wide range of devices.

\item Serialization and export. \vp provides three output mechanisms for different purposes. Users
can export snapshots of the current plot to a Portable
Document Format (PDF) file using a combination of the \texttt{Canvas2Svg}\footnote{\url{https://github.com/gliffy/canvas2svg}} and
\texttt{SVG-to-PDFKit}\footnote{\url{https://github.com/alafr/SVG-to-PDFKit}} libraries, providing
a convenient format for documenting, sharing, or printing the resulting visibility plot and schedule.
The application state
can be saved to a compressed \texttt{.visplot} file, which is the primary format for
preserving and subsequently restoring a session. This file contains
the generated schedule, target metadata, observing configuration, customized constraints,
and other application settings required to resume work, and
can be reloaded on another device running \vp.
Finally, for supported telescopes, \vp can export a TCS catalogue containing the coordinates of all targets.
This allows the catalogue to be loaded directly into the TCS, where targets can then be selected and pointed to without
manual entry of their coordinates.
\end{itemize}

\subsection{Observatory integration}\label{software:integration}
Adapting \vp to a new observatory begins by defining a telescope configuration
in JavaScript. At a minimum, this specifies the observatory name, geographic coordinates,
elevation above sea level, and IANA time-zone identifier. Additional parameters can be
supplied to model site- or instrument-specific behaviour, including instrument definitions
(field of view, orientation, etc.) and a background image for the plot.
Mechanical telescope limits may also be supplied by the user. These could be derived either
analytically (e.g. from mount geometry) or empirically through measurements of the
telescope's accessible sky, such as collision maps. Once available, these limits can
be represented as simple analytical expressions (e.g. declination-dependent altitude limits)
or through arbitrary JavaScript functions implementing lookup tables, spline interpolation,
or other numerical models (as discussed in \secref{model:visibility} below), allowing
complex mount- and instrument-dependent restrictions to be modelled.
Similarly, integration of an all-sky camera requires an astrometric calibration relating
image pixels to altitude--azimuth coordinates. This calibration is observatory-specific
and must be determined independently using standard calibration procedures (e.g. \citealp{barghini2019}),
after which the resulting parameters can be incorporated directly into the telescope configuration.
The existing telescope configurations provide practical starting points for new deployments,
encompassing both altazimuth and equatorial telescopes, multiple instrument configurations,
and a variety of observatory-specific mechanical constraints.

\begin{table*}[h!]
\caption{Examples of input formats accepted by \vp.}
\label{table:input}
\centering
\begin{tabular}{p{9.4cm}p{8.1cm}}
\hline \hline
\noalign{\vskip 2pt}
Input format  & Comments \\
\hline
\noalign{\smallskip}
18 36 56.336 +38 47 01.28 & RA and Dec in sexagesimal representation\\
279.2347333 38.7836889 & RA and Dec in decimal degrees\\
EQPsc 23 34 34 -01 19 36 & Name, RA and Dec \\
EQPsc 23:34:34.70 -01:19:36.01 & Colon-separated RA and Dec are also accepted \\
HD84937 09 46 12.06/0.373 13 59 17.44/-0.774 1950 & Proper motions (arcsec/yr) appended to RA and Dec \\
Targ 12:0:0 35:0:0 2000 300 54-321 2.0 ToO ALFOSC 15 2 & \hspace{-1pt}All parameters specified explicitly\\
EQPsc 23 34 34.70 -01 19 36.01 2000 1800 54-321 UTC[22-24] & Force scheduling inside the specified UTC range \\
EQPsc 23:34:34.70 -01:19:36.01 2000 * 54-321 UTC[20-20:30] & Force filling the entire UTC range with this target \\
Targ 12:0:0 35:0:0 2000 * 54-321 AM1.4,DARK & Observe this target for as long as it fulfils its constraints \\
Jupiter & \vp calculates the ephemerides of the planets/Moon\\
Aldebaran & RA, Dec, and proper motions retrieved from SIMBAD \\
"alf Tau" & Same as above \\
"2026 FG4" & Ephemerides retrieved from JPL\\
Offline LST[23-3.5] & No observations scheduled in this UTC or LST range \\
\hline
\end{tabular}
\tablefoot{The available fields for each target are: name,
RA (+proper motion in RA), Dec (+ proper motion in Dec), epoch, duration, project identifier,
comma-separated list of constraints, type, instrument, sky PA, and priority.}
\end{table*}

In many observatory environments, observations are organized as observing
blocks (OBs) that encapsulate target information, constraints, and execution
parameters. \vp integrates naturally with such systems.
Observing blocks can be preloaded via a Hypertext Transfer Protocol (HTTP) POST request
containing JavaScript Object Notation (JSON)-encoded
OB information, enabling automated preparation of nightly
schedules triggered by external systems. Alternatively, local deployments may populate the
application state directly through PHP session variables. These interfaces allow observability
analysis and preliminary scheduling of complete nightly OB queues with minimal user interaction.
The documentation\footnote{\url{https://visplot.readthedocs.io/en/latest/observatory_integration.html}}
includes concrete examples of the required POST payload and
session structure allowing external systems to pre-populate \vp fields
programmatically.

Although \vp is fundamentally a client-side application, a minimal PHP
abstraction layer deployed on the hosting server supplements the primary client-side architecture.
This thin backend serves two functions required to bridge the browser sandbox with external systems.
First, it parses incoming session data transmitted
by external automated systems, such as OB queues, allowing the 
application to initialize state and pre-populate the target list dynamically. 
Second, it acts as a reverse proxy for harvesting data from external infrastructure:
local all-sky cameras, real-time TCS coordinates, and external JPL ephemeris servers. 
This proxy mechanism is strictly necessary to circumvent browser-enforced Cross-Origin 
Resource Sharing (CORS) restrictions, which would otherwise block client-side 
JavaScript from establishing direct connections to these decoupled network resources.

\subsection{Distribution and documentation}\label{software:distribution}
A public deployment of
\vp\footnote{\url{https://www.visplot.com}}
allows unrestricted, immediate use without installation, serving as both a
reference implementation and a convenient entry point for rapid planning, proposal
preparation, and educational use. Alternatively, containerized deployment
(via \texttt{Docker}) enables reproducible, self-contained installations that simplify
integration with observatory systems and allow site-specific customization while
maintaining the same browser-based interface.
The code is publicly available on GitHub\footnote{\url{https://github.com/egafton/visplot}},
with version-tagged releases ensuring reproducibility and supporting long-term
operational stability.
The user interface is documented directly in the web interface via a `Help'
button, and more extensively through a dedicated documentation site hosted on
Read the Docs\footnote{\url{https://visplot.readthedocs.io}}.
This documentation includes tutorials, configuration examples, installation
instructions, and guidance on reporting issues or requesting new features.

\section{Conceptual model and implementation}\label{sec:model}
\vp occupies the operational space between static visibility calculators and
large-scale survey schedulers, and addresses a very common operational use case:
a finite list of known targets to be observed during a single night, most of which
are expected to be executed if conditions allow. In this context, the
software must address two closely related problems. The first is to determine
the observability windows of each target (including non-sidereal ones) under the relevant astronomical and
operational constraints. The second is to construct a feasible 
and efficient observing schedule within those windows, while remaining flexible to
real-time adjustments by the observer. This approach reflects common operational
practice at small and medium observatories, where schedules are frequently adjusted
in response to changing weather, technical issues, or override triggers,
and human judgement remains central to decision-making.

\subsection{Visibility and constraint model}\label{model:visibility}
At its core, \vp calculates the altitudes of astronomical targets on an
equidistant time grid spanning the observing night.
Users provide the coordinates and constraints for a list of $N$
objects (using various formats, as shown in \tabref{table:input}), 
and the software generates a plot showing their `observed altitudes'
and corresponding observability regions, as defined below. 

Targets are specified by their mean ICRS, J2000 or B1950 right ascension (RA)
and declination (Dec), and (optionally) proper motions at a specified epoch. 
Parallax corrections are not applied to sidereal targets, for which the effect
is negligible for altitude-chart generation. For non-sidereal objects,
topocentric astrometric coordinates are obtained from JPL Horizons, thereby
incorporating the geometric observer-location effects.
Altitudes are computed over the full time domain, including periods when an
object is below the horizon, although negative values are not displayed on the
plot.
`Observed' (as opposed to mean or apparent) altitudes include the full
calculation of light deflection, annual aberration, precession and nutation,
Earth rotation, diurnal aberration, and estimated atmospheric refraction.
Apart from target altitudes, the software also calculates sunset, sunrise and
twilight times, as well as the moon altitude, illumination, and angular distance
to each target throughout the night.
The sunset and sunrise times define the limits of the night itself (and
therefore of the plot). They are computed from topocentric solar ephemerides by
first determining the time of solar transit and then solving for the instant at
which the refracted upper limb of the Sun coincides with the local horizon.
Nautical and astronomical twilight times are computed in a similar manner but
using altitudes of $-12^\circ$ and $-18^\circ$, respectively, and without
refraction or solar-radius corrections.

Following the ephemeris calculations, \vp performs a pass through the time grid
to determine the observability windows of each target. For every point, the
software evaluates the target altitude together with all applicable target, 
telescope, and instrument constraints. Points that satisfy all constraints in
the `observability grid' (\citealp{morris2018}) are classified as observable,
and contiguous sequences of such points are merged into observability windows.

Let the observing night be represented by a continuous time domain bounded
by the sunset and sunrise times,
\begin{equation}\label{eq:night}
T = \left[ t_{\rm set}, t_{\rm rise} \right].
\end{equation}
For each target $i\in\left\{1,\dots,N\right\}$, we define the observability region
as 
\begin{equation}\label{eq:region}
W_i \subseteq T.
\end{equation}
Because various constraints may split observability into multiple segments, we
represent $W_i$ as a finite union of disjoint closed intervals representing
individual observability windows,
\begin{equation}\label{eq:segments}
W_i = \bigcup_{k=1}^{K_i} \left[ s_{ik}, e_{ik}\right],
\end{equation}
where $K_i \ge 0$ is the number of valid observability windows, $s_{ik}$ and
$e_{ik}$ are the start and end times, respectively, of window $k$, and the
windows are ordered and non-overlapping, $e_{ik} < s_{i,{k+1}}$.
If $K_i$=0, target $i$ is not observable that night.

\begin{table*}[!h]
\caption{Summary of target-specific constraints used by \vp to define the feasibility windows for each observation.}
\label{table:constraints}
\centering
\begin{tabular}{lp{11cm}p{4cm}}
\hline \hline
\noalign{\vskip 2pt}
Constraint & Description & Examples \\
\hline
\noalign{\smallskip}
Airmass & Hard cut-off to $[X_{\rm min}, X_{\rm max}]$. Can specify both $X_{\rm min},X_{\rm max}$ or just $X_{\rm max}$ (in which case $X_{\rm min}=1$) 
        & 2.0 \newline AM1.5 \newline AM[1.2-1.6] \\
\noalign{\vskip 2pt}
Moon distance & Hard cut-off to $[D_{\rm min}, D_{\rm max}]$. Can specify both $D_{\rm min},D_{\rm max}$ or just $D_{\rm min}$ (in which case $D_{\rm max}=180$\textdegree) 
              & MOON40 \newline MOON[40-180] \\
\noalign{\vskip 2pt}
Time interval & Hard cut-off to given UTC, LST or HA interval. Must specify both limits 
          & UTC[23:30-02:00] \newline LST[11-13] \newline HA[0-2] \\
\noalign{\vskip 2pt}
Twilight mode & Schedule only within nautical twilights (NT), astronomical twilights (AT), dark time (DARK), or a combination thereof. Overrides global settings 
              & NT \newline AT+DARK \newline DARK \\
\noalign{\vskip 2pt}
Offline & No observations can be scheduled within the given UTC or LST range 
        & Offline UTC[23:45-1:00] \newline Offline LST[10-11] \\
\hline
\end{tabular}
\end{table*}

Target-specific constraints are listed in \tabref{table:constraints}, and include
airmass and moon distance ranges, explicit coordinated universal
time (UTC), local sidereal time (LST) or hour angle (HA) observing windows, 
and a twilight mode flag. The user can also specify global `offline' intervals, defined as UTC or
LST ranges during which no observations can be scheduled
(e.g., because of weather conditions, technical downtime, or time reserved for
another user).
These filters can be combined to accommodate arbitrarily complex 
observational requirements.

Telescope and instrument constraints represent hard operational limits, such as
minimum and maximum telescope altitudes, collision avoidance regions, or
instrument-specific restrictions, and are defined globally as part of the
telescope configuration (\secref{software:integration}).
Altitude constraints depend strongly on the telescope mount geometry and
mechanical limitations. Altazimuth mounts typically impose a fixed minimum
altitude (which may be further restricted by vignetting when observing with a
closed lower hatch) $a_{\rm min}$, and a maximum altitude limit $a_{\rm max}$
that reflects tracking limitations near zenith caused by finite azimuth drive speeds.
For instance, in case of the Nordic Optical Telescope (NOT),
\begin{subequations}
\begin{align}
a_{\rm max} &= 88^\circ, \\
a_{\rm min} &= 6^\circ.
\end{align}
\end{subequations}
Equatorial mounts, by contrast, often exhibit additional minimum altitude limits
expressed as a function of declination $\delta$ or azimuth $A$. Often, these
are closed-form empirical functions, such as in the case of the Isaac
Newton Telescope (INT), where
\begin{subequations}
\begin{alignat}{3}
a_{\rm min} &= \rlap{$\arcsin(\sin\phi\cdot\sin\delta),$} & & & +90^\circ \ge \delta > 45^\circ 18\arcmin 02\arcsec, \\
a_{\rm min} &= 20^\circ,     & \hphantom{\qquad\qquad\quad} & & 45^\circ 18\arcmin 02\arcsec \ge \delta \ge -30^\circ 09\arcmin 30\arcsec, \\
a_{\rm min} &= 90^\circ,                                 &  & & -30^\circ 09\arcmin 30\arcsec > \delta \ge -90^\circ, \label{eq:intlowdec}
\end{alignat}
\end{subequations}
where $\phi\approx +28.76^\circ$ is the telescope latitude, and
\eqreft{eq:intlowdec} reflects that targets below that declination cannot
be observed at the INT.

In other cases, $a_{\rm min}$ can only be described through a more complex
collision map, as in the case of the Harlan J. Smith Telescope, whose
cross-axis equatorial yoke mount operates within a dome with multiple
configuration-dependent mechanical obstructions\footnote{\url{https://mcdonald.utexas.edu/media/165/download?inline}}.
These limits are generally governed by mount design, railings, cables, ladders,
and other structures, and are specific to each telescope and, in some cases, are further
restricted by the instrument in use. For equatorial mounts that require meridian
flips, the limits also depend on whether the telescope is in a tube-west or a
tube-east configuration.

\vp can handle all these cases through custom functions that describe
$a_{\rm min}(\delta)$ or $a_{\rm min}(A)$ via analytical expressions, lookup
tables, or spline interpolation. This flexible approach allows a wide range of
telescope and instrument configurations to be modelled consistently.
Custom limits can be added to a local deployment of the software
and then integrated into the main codebase via pull requests.
Alternatively, they can be reported via the project issue tracker or communicated
directly to the authors, who can implement the corresponding modifications into the
main branch.\footnote{More details are available at \url{https://visplot.readthedocs.io/en/latest/issues.html\#requesting-new-features}}

The observability windows for each target form the fundamental building blocks for
subsequent scheduling.

\subsection{Scheduling model}\label{model:scheduling}
Astronomical scheduling is an NP-hard problem (\citealp{gomez2003}), meaning
that the time required to find the optimal solution can increase very rapidly as the 
number of targets and constraints grows. In practice, this makes exact optimization 
computationally expensive, so heuristic approaches are commonly used.
The scheduling problem has been extensively studied in the context of large
surveys, robotic telescopes, and long-term observing campaigns (e.g., \citealp{garcia2017}). In such cases,
thousands of targets must be scheduled over an extended period of time (ranging from
months to years) based on multiple criteria consolidated into a merit function
(e.g., \citealp{steele1997,denny2004,bierwirth2010}) using various global
optimization techniques (e.g., \citealp{naghib2019,bellm2019}). Many approaches
have been analysed and compared in the literature (e.g., \citealp{mora2010,colome2012,fraser2012,solar2016})
and frameworks for their relative assessment have also been published
(\citealp{xie2024}).

\begin{table*}[!h]
\caption{Overview of the heuristic weighting criteria and default weights used
to calculate the total scheduling score $M_i$ for each target $i$
(\eqrefp{eq:score}).}
\label{table:weights}
\centering
\begin{tabular}{p{1.5cm}p{8.8cm}lll}
\hline \hline
\noalign{\vskip 2pt}
Criterion & Description & Formula $X_{k,i}$ & Default $W_k$ & Score\\
\hline
\noalign{\smallskip}
Mandatory override & Bypasses heuristic scoring; target is pre-scheduled to fill its first available
                     feasibility window before the main algorithm runs. & N/A & N/A & N/A \\
\noalign{\vskip 2pt}
Priority & Based on the user-assigned priority $p_{i}$; ensures that the highest-value scientific targets
           are favoured in the scheduling sequence. & $P_i=p_i/p_{\rm max}$ & $W_p=2$ & $W_p \times P_i$\\
\noalign{\vskip 2pt}
Urgency & Based on the time remaining until the target reaches its latest observable limit $t_i^{\rm max}$ (\eqrefp{eq:latest});
          prioritizes observations nearing the end of their feasibility window. &  $U_i=1/(t_i^{\rm max} - \tau + 1)$ & $W_u=1$ & $W_u \times U_i$ \\
\noalign{\vskip 2pt}
Altitude & Based on the target's instantaneous altitude $a_i(\tau)$; favours observations at lower airmass,
           maximizing scientific data quality and signal-to-noise ratio. & $A_i=\sin a_{i}(\tau)$ & $W_a=1$ & $W_a \times A_i$ \\
\noalign{\vskip 2pt}
Slewing time & Based on the angular separation $\Delta\theta$ (great-circle distance) between
               consecutive targets; minimizes telescope overhead. &  $S_i=1-\Delta\theta_{i,j}/\pi$ & $W_s=1$ & $W_s \times S_i$ \\
\hline
\end{tabular}
\end{table*}

The observability windows defined in \secref{model:visibility} above establish 
when each target may be observed. Scheduling constitutes an additional step in which 
these windows are combined with target execution times and user-defined priorities 
to construct a feasible observing sequence for the night. \vp approaches this problem 
in two stages (a strategy common in the literature, e.g. \citealp{sessoms2009}), 
inspired by how observers would often 
plan their observations using static charts (e.g., \staralt) and manual scheduling.
First, targets marked as mandatory (in the way described in \secref{workflow:planning} below) are pre-allocated to guarantee their 
execution; second, the remaining targets are selected iteratively from the pool of 
currently feasible observations according to a scheduling algorithm, in order to fill
the gaps left by the pre-allocation.
The proposed schedules can be manually modified even as the observing
night is in progress, allowing observers to adjust them in order to
accommodate sudden target-of-opportunity (ToO) observations, technical interruptions,
adverse weather, or other unanticipated circumstances.
The quantities defined below form the mathematical basis for evaluating and 
prioritizing candidate observations during scheduling.

Assuming that each target $i$ has an execution duration
\begin{equation}\label{eq:duration}
d_i > 0,
\end{equation}
a start time $t \in T$ is feasible if the full execution fits
inside one of its observability windows $W_i$ (\eqrefp{eq:segments}). Therefore, the feasible start-time set is
\begin{equation}\label{eq:feasible}
F_i = \bigcup_{k=1}^{K_i} \left[ s_{ik}, e_{ik}-d_i \right].
\end{equation}
If an interval $k$ is shorter than the required duration, $e_{ik}-s_{ik} < d_i$,
that interval does not contribute to $F_i$.

We can then express the latest feasible starting time, $t_i^{\rm max}=\sup F_i$,
using the fact that $F_i$ is a finite union of closed intervals, as:
\begin{equation}\label{eq:latest}
t_i^{\rm max} = \max_{k \in \left\{1,\dots,K_i\right\}} \left( e_{ik} - d_i \right),
\end{equation}
which falls inside the last interval that satisfies $e_{ik}-s_{ik} \ge d_i$.
If none of them do, then target $i$ is not schedulable for the night.

During its first decade, until the release of version 5.0, \vp employed a simple greedy algorithm for the second
phase, in which targets were ordered strictly by their latest observable time
$t_i^{\rm max}$, implementing an `earliest deadline first' strategy.
This approach is mathematically equivalent to the classical unweighted
activity-selection problem (\citealp{cormen2009}), proven to yield an optimal
solution when the objective is to maximize the total number of non-overlapping
activities. While robust, this model does not account for user-defined priorities
or operational overheads.

The current implementation generalizes this second phase by allowing different
scheduling algorithms to be applied to the remaining targets. The default option
is a weighted greedy heuristic, in which candidate targets $i$ are dynamically
ranked at each time step $\tau$ using a merit function
\begin{equation}\label{eq:score}
M_i(\tau) = \sum_k W_k \times X_{k,i}(\tau),
\end{equation}
where $W_k$ are user-configurable weights (which can be positive, negative, or zero),
and the terms $X_{k}$ represent
the criteria listed in Table~\ref{table:weights}: user-defined scientific
priority $P_i$, urgency (proximity to $t_i^{\rm max}$) $U_i$,
instantaneous altitude $A_i$, and the great-circle slew distance from the
previous telescope position $S_i$.
All $X_{k}$ terms are formulated to yield values within the interval $[0,1]$.
Targets for which $\tau\notin F_i$ are not considered for scheduling at time step $\tau$.
The target that maximizes the multi-objective function $M_i(\tau)$ is selected, the time 
pointer $\tau$ is advanced by $d_i$, and the process is repeated for the remaining 
target pool. Compared to the earlier activity-selection approach, this weighted,
constraint-aware scheme produces schedules that remain feasible while better
reflecting user-defined priorities and operational considerations.

In addition to the default heuristic scheduling method, we also implemented a flexible-placement
beam search (e.g., \citealt{russell2010}) that explores multiple candidate
observing sequences in parallel.
In this method, observations are not assigned fixed start times 
during the search. Instead, each target is represented by one or more allowed time 
windows, and each candidate schedule maintains, for every target, an earliest 
and latest feasible start time. When a new target is inserted into a sequence, the 
algorithm considers all possible insertion points, and for each of them it
enforces temporal consistency by propagating constraints forward and backward along the sequence.
This propagation step incrementally tightens the feasible start-time ranges of all 
affected targets and immediately rejects sequences in which any interval becomes infeasible.
Operating with flexible time intervals rather than fixed timestamps allows
the scheduler to defer placement decisions until sufficient context 
is available. This significantly reduces the risk of early choices blocking later 
feasible configurations, a common limitation of purely greedy approaches.
At each expansion step, the beam is pruned to retain only the highest-scoring
partial schedules (typically of order $\sim 10$). These states are ranked using
the merit function in \eqreft{eq:score}, applied cumulatively to all targets
included in the sequence so far.
The method follows standard principles of constraint-based temporal reasoning
(e.g., \citealt{dechter2003}).

More generally, the \vp scheduler is implemented as a modular component that
operates on a set of targets and returns an ordered sequence, while all constraint
evaluation and feasibility checks are handled independently. This separation allows
alternative algorithms (including exact solvers) to be incorporated as drop-in replacements without 
modifying the surrounding infrastructure. In this sense, \vp is not defined by 
a specific scheduling strategy, but by the broader framework that integrates 
constraint evaluation, scheduling, and interactive visualization within a 
unified operational environment. As of version 5.5, both scheduling methods presented above
remain available, and users can select either method at run time. For typical target list sizes
(below 50), we did not measure a significant difference in execution time between the two approaches on 
modern hardware. The beam-search algorithm generally produces schedules that are 
closer to the global optimum, particularly in cases with competing priorities and 
tightly constrained observability windows. The weighted greedy heuristic, however, 
remains useful because its behaviour is transparent and deterministic: scheduling 
decisions can be directly traced to the merit function evaluated at each step, making 
the resulting schedules easier to interpret and predict. The two methods therefore 
represent a trade-off between global optimization and algorithmic simplicity, 
and are provided as complementary options within the framework.\footnote{A 
more in-depth comparison between the two algorithms is available at 
\url{https://visplot.readthedocs.io/en/latest/tutorial.html\#when-the-schedulers-disagree-a-deeper-comparison}}

\section{Operational workflows}\label{sec:workflow}
\subsection{Nightly planning and proposal preparation}\label{workflow:planning}
\figref{fig:overview} shows an overview of the \vp web interface after a
candidate schedule has been generated, illustrating how visibility charts and
scheduling information are presented within a single interactive workspace.
In this example we are using a selection of heterogeneous targets, as
detailed in \secref{workflow:example} below.

The user first specifies the observing date (\panelref{A1}) and (if necessary)
can change the telescope and default settings in the configuration panel (\panelref{A2}).
These settings include the default epoch, total duration (exposure time + overheads), project identifier,
maximum airmass, type of observation, instrument, and plotting colours.
All settings are stored locally in the browser, so they only need to be
configured during the first use of \vp on a given device.

The target list is then provided (\panelref{A4}), which may contain
up to thirteen fields: name, RA and Dec, proper motions in RA and Dec, epoch, 
duration, project identifier, observing constraints (airmass, UT, or LST), observation type,
instrument, sky position angle, and priority (as shown in \tabref{table:input}).
The only mandatory fields are either the name alone (in 
which case \vp will automatically retrieve the RA, Dec, and 
proper motions from SIMBAD or JPL Horizons, if available), or the RA and Dec.
Missing fields are automatically filled with the configured defaults.
An asterisk in the duration field marks the target as mandatory, and ensures that
it will be scheduled with absolute priority for as long as it fulfils all of its constraints,
as discussed in \secref{model:scheduling}.
UTC or LST time windows in which no observations can be scheduled (either due to
bad weather, or because e.g. the night is shared with other observing
programmes) can be specified using the \texttt{Offline} syntax shown in
\tabref{table:input}. An example of how \vp renders such a window is shown in
\panelref{B1}, where observations are scheduled around the UTC 2--3
range marked as offline and rendered as a grey shaded area.

For solar system objects, the assumption of fixed celestial coordinates 
is insufficient. \vp therefore supports both internally computed and externally
retrieved ephemerides. For major bodies such as the Moon and planets, ephemerides 
are computed directly within the software based on the algorithms of \citet{meeus1998},
providing near-instantaneous results with an accuracy of order $\sim$1 arcmin or better.
For minor bodies (including comets and asteroids), ephemerides are retrieved
from JPL Horizons. Upon entering a valid
designation or Horizons ID\footnote{\url{https://ssd-api.jpl.nasa.gov/doc/horizons.html}},
the software retrieves the object's RA and Dec at a configurable sampling interval (30 min by default)
throughout the specified night. This default cadence provides sufficient accuracy 
for the vast majority of non-sidereal targets, while limiting network traffic and 
unnecessary load on the Horizons service. For local deployments, the sampling interval 
can be adjusted to accommodate applications requiring finer temporal resolution.
The retrieved ephemerides are then linearly interpolated on to the internal time grid,
allowing the constraint engine to treat non-sidereal targets with the same
mathematical rigour as sidereal ones, dynamically updating lunar separation and
airmass as the object moves across the sky.
Following the target resolution via JPL Horizons, the target list (\panelref{A4}) will show
the RA and Dec at the beginning of the night (i.e., at sunset) for non-sidereal targets,
while \vp will internally store the interpolated ephemerides throughout the night. 
If the state of the application is saved and reloaded (\panelref{A6}), the interpolated
RA and Dec are preserved. However, if the list of targets is copy-pasted in the text box,
all RA and Dec are interpreted as sidereal, and the visibility plots for non-sidereal
targets may be inaccurate. This is a current limitation of the software that will be addressed
in a future update.

\vp also provides four predefined lists of calibration targets
that can be accessed by 
clicking on the corresponding button in the interface (\panelref{A3}). The first two are
deep blank-field catalogues for the northern and southern hemisphere (\citealp{jimenez2012}), while the other two are spectrophotometric
standard stars compiled from various catalogues, taken from the Isaac Newton Group of Telescopes (ING) 
website\footnote{\url{https://www.ing.iac.es/Astronomy/observing/manuals/html\_manuals/ tech\_notes/tn065-100/workflux.html}}
for the northern hemisphere, and from the European Southern Observatory (ESO) 
website\footnote{\url{https://www.eso.org/sci/observing/tools/standards/spectra/stanlis.html}}
for the southern hemisphere.

After the list of targets is filled in, the user can plot their altitudes by pressing 
the corresponding button (\panelref{A5}).
\vp then computes the observability of each target over the course of the
given night, using the methodology described in \secref{model:visibility}, and
presents interactive plots for evaluation (\panelref{B1}). In the altitude
plots, time intervals that satisfy all the constraints are shown distinctly
(solid black line) from those that do not (faint grey line), allowing users to
visually identify viable observing windows.

Based on the computed observability windows, \vp can generate a candidate observing
schedule as described in \secref{model:scheduling} when the user presses
the corresponding button (\panelref{A5}). Users may reorder the
targets by drag-and-dropping them in the list (\panelref{B3}) in order to account for more
complex operational constraints. This interactive loop supports common
operational practices in which observing plans are refined incrementally or
on-the-fly rather than being fixed in advance.

After the night is scheduled, a quantitative summary of the plan is provided
in the log (\panelref{A7}), including the total night length (with dark and twilight components),
the fraction of time allocated to observations, any predefined offline intervals,
and the remaining unscheduled time. In addition, the observing time is broken down
by project identifier, allowing users to verify how time is distributed across programmes.
These diagnostics provide an immediate quantitative assessment of schedule completeness and
efficiency.

Beyond nightly planning, the same workflow can be used during proposal preparation. 
Because observability analysis and scheduling can be performed for arbitrary dates, investigators 
can explore different observing scenarios by varying target selections, observing constraints, 
or telescope facilities. This allows rapid assessment of seasonal observability, 
lunar conditions, and expected schedule utilization before a proposal is submitted, 
helping to verify that the requested observing time is realistic and compatible with 
the operational constraints of the intended facility.

The application state can be saved to a file using the controls
in \panelref{A6}, and subsequently reloaded on any device. In the context
of modern remote and geographically distributed observing, this feature allows a
principal investigator to prepare an intricate nightly schedule at their home 
institution in advance, and digitally share it with a telescope 
operator or visiting observer at the remote facility. Rather than relying on 
static charts or text-based lists, the operator receives a fully interactive
`mission plan' that can be rapidly adjusted if local conditions change.

\subsection{Scheduling example for a representative night}\label{workflow:example}
To illustrate the behaviour of the scheduling engine under realistic conditions,
we consider a typical observing night with a heterogeneous set of targets 
and constraints. The input list (\panelref{A4})
includes standard calibration fields from the predefined target list (\panelref{A3}), science 
targets spanning a wide range in RA and Dec, and a few 
non-sidereal objects (Jupiter, comet C/2022 QE78, and asteroid 2026 CO).
In addition to standard altitude and airmass constraints, the input 
incorporates hour angle, lunar distance, and time-window constraints (both in 
UTC and LST), as well as a fixed offline interval representing scheduled 
technical downtime. The constraints are specified and applied as described in \secref{model:visibility}.

A visualization of the resulting schedule is shown in \panelref{B}.
The altitude curves encode constraint satisfaction explicitly: dark segments
indicate intervals where all constraints are fulfilled, while light grey segments
denote times when at least one constraint is violated. The scheduler operates 
exclusively on the valid (dark) intervals, ensuring that all scheduled observations
satisfy the full set of constraints.

The scheduling engine produces within a fraction of a second a contiguous, non-overlapping 
sequence of observations  that satisfies all specified constraints, including 
the dynamically evaluated positions of non-sidereal targets.
Time-critical constraints are correctly enforced: targets meant to fill their 
entire window (drawn in orange) are scheduled within their respective 
intervals, while the offline period is respected as an exclusion zone within the 
schedule.
Target priority is encoded by colour (from highest to lowest: magenta, blue, green).
The resulting schedule reflects the heuristic's multi-objective nature:
targets are generally scheduled near culmination to maximize
altitude (and thus data quality), unless this conflicts with
priority, urgency, or slew-minimization constraints (\secref{model:scheduling}).
For this example, we used the flexible-placement beam search algorithm (\secref{model:scheduling})
with default heuristic weights (\tabref{table:weights}). Both the greedy
and beam search algorithms schedule all targets, but the greedy
prioritizes targets 10 over 16 and 18 over 13, leading to higher airmasses.
In practice, this trade-off is operational: observers may prefer higher airmass for high-priority
targets to reduce the risks of postponement, and the final decision is left to their judgement.

\begin{figure}[t!]
\begin{center}
\includegraphics[width=\columnwidth]{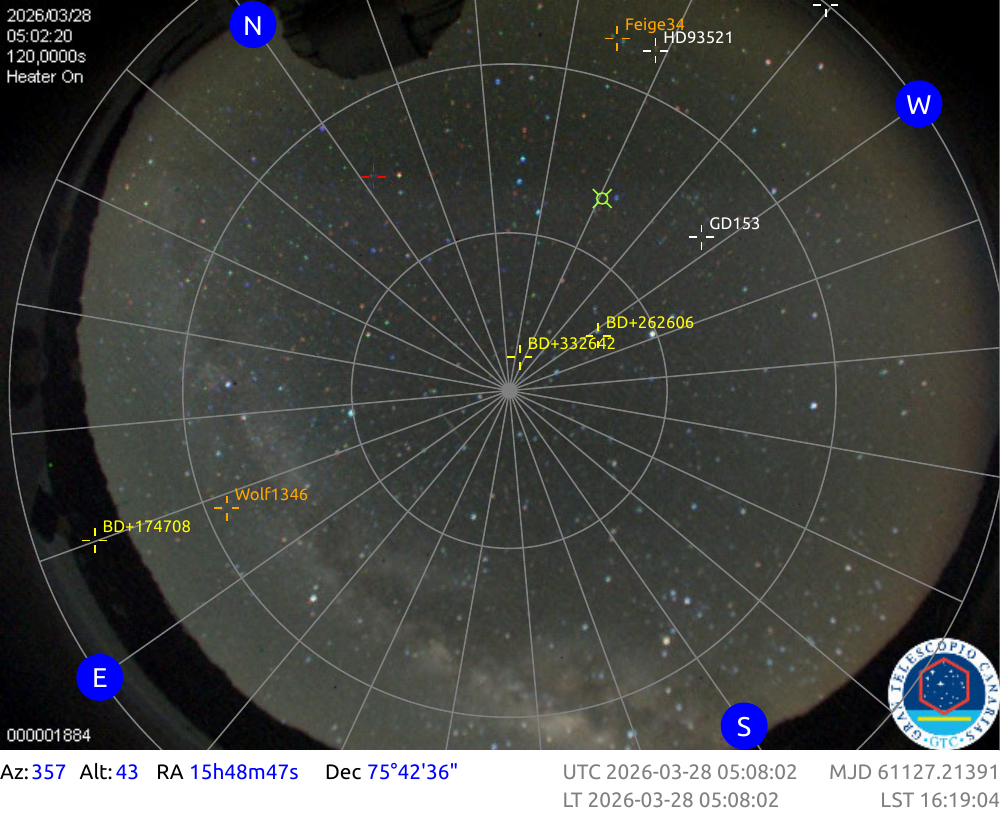}
\end{center}
\caption{Gran Telescopio Canarias (GTC) all-sky camera image showing the position of targets on-sky. 
Apart from target overlays (labelled with the object names), the current TCS pointing 
(if available) is indicated by a light green crosshair, while the pointing position 
under the user's mouse cursor is shown with a red crosshair. The coordinates of
the latter (azimuth, altitude, RA, Dec) are displayed in the lower-left corner.
In the bottom-right corner, the current UTC (both ISO and MJD), local time (LT), and local 
sidereal time (LST) are updated every 500 ms.}
\label{fig:skycam}
\end{figure}

A quantitative summary of the example schedule (\panelref{A7}) 
further illustrates its efficiency. For a 
total night length of 11\,h 24\,min, including 8\,h 59\,min of dark time, the scheduler allocates 
10\,h 24\,min (91\%) to observations, while preserving the 1h (9\%) offline 
interval specified in the input. The remaining unscheduled time is negligible 
(7 s, <0.02\%), indicating that the available observing window is effectively 
filled. In this example, the total duration of execution was chosen to closely 
match the available night length. Under such conditions, and in the presence of 
heterogeneous constraints, the number of possible schedules that fully utilize 
the night is limited: early scheduling choices can easily lead to conflicts or 
unusable gaps later on.

In this example, the observing night also includes a transition from standard time
to daylight saving time, which is explicitly reflected in the time axis. 
The local time scale exhibits the corresponding one-hour shift (from WET to WEST), 
while UTC and LST remain continuous. This behaviour is not merely cosmetic: correct 
handling of time system transitions is essential for operational planning, particularly 
when schedules are communicated in local time or executed across system boundaries. 

The results shown here demonstrate that the heuristic approach yields schedules 
that are well balanced in terms of constraint satisfaction, data quality, and 
slewing overhead, while remaining sufficiently fast to support interactive 
use during both preparation and execution of an observing night.

\subsection{Real-time execution}\label{workflow:realtime}
During an observing night, a vertical line drawn in red (as it appears shortly
before 3:00 UTC in \panelref{B1}) and moving right as the night progresses
indicates the current time, assisting observers in real-time decision-making.
\vp also supports the integration of an observatory-specific all-sky camera
image (\panelref{A6}), like the one shown in \figref{fig:skycam}, allowing the current target 
positions to be plotted on sky. This is especially useful during partially
cloudy nights, when it enables the observers to dynamically select visible targets. 

\begin{figure}[t!]
\begin{center}
\includegraphics[width=\columnwidth]{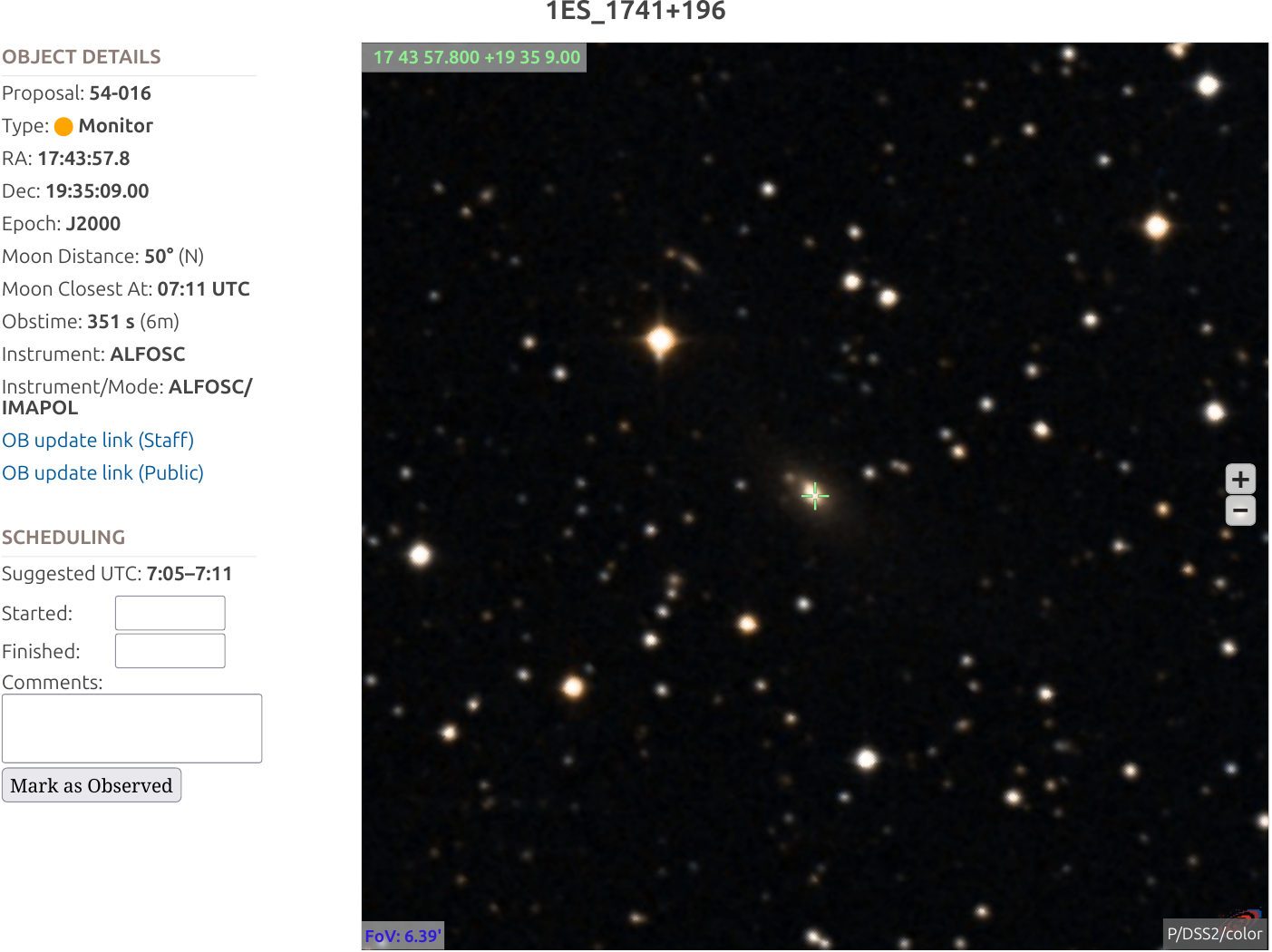}
\end{center}\caption[]{
Finding chart of \object{1RXS J174357.3+193505} generated by \vp,
accessible by selecting a target index directly from the visibility plot.}
\label{fig:finding_chart}
\end{figure}

Finding charts are fully integrated into the interactive workflow of \vp through
Aladin Lite v3. The surveys used for display are the Digitized Sky Survey 2 (DSS2; \citealp{lasker1996}) for optical instruments,
and the Two Micron All Sky Survey (2MASS; \citealp{skrutskie06}) for infrared instruments.
If no instruments are defined for the selected telescope (as discussed in \secref{software:integration}) the optical survey is used by default.
The finding chart can be displayed by
clicking on the target index directly in the visibility plot (\panelref{B1}), 
allowing rapid visual identification of the field of view (\figref{fig:finding_chart} illustrates one such example).
The field of view of the finding chart matches that of the selected instrument, if
defined in the telescope configuration (\secref{software:integration});
otherwise, a configurable fallback field of view is used (6 arcmin by default).
Target-specific sky position angles are also supported, in order to have the
finding chart match the orientation of the image expected on the CCD.
The finding chart supports interactive zooming and panning, which is particularly
useful for identifying targets in cases of modest telescope pointing accuracy.
The finding chart is accompanied by object metadata, including name and celestial
coordinates, angular distance from the Moon, and the user-defined duration of observation.
In cases where the OB framework is integrated with \vp (such as at the NOT),
the panel also includes backlinks to the OB queue, allowing the observer to quickly access the 
OB and record comments (e.g., regarding the observing conditions, data quality, etc.)
after the observation has been carried out.
When a nightly schedule has been generated, the scheduling panel also displays
the suggested observing interval and provides a \texttt{Mark as Observed} button.
A target such marked will then be shown in green and will remain
unaffected by subsequent changes in the schedule.

In general, there are a number of situations where the observer would
need to modify the schedule dynamically. These include changing weather 
conditions that degrade seeing or sky transparency, or the sudden arrival of 
high-priority ToO triggers. In the era of time-domain and multi-messenger
astronomy, rapid response to transient events (such as gamma-ray bursts or
gravitational wave alerts) is critical. \vp allows the observer to update the
target list (\panelref{A4}) at any time by adding new targets, modifying
constraints or priorities, or removing targets, and to regenerate the schedule
on-the-fly (\panelref{A5}). During rescheduling, targets already marked as
`Observed' are left unchanged, while the remaining targets (including any newly
added ones) have their observability recomputed and are rescheduled starting
from the current time rather than from the beginning of the night.

\section{Adoption and impact}\label{sec:adoption}
\subsection{Operational use}
Since its initial deployment at the NOT in 2016 and its
subsequent public release as open-source software starting with version 2.0 in 2020,
\vp has been used at multiple observatories worldwide, including Roque de
los Muchachos, Teide, Calar Alto, Mauna Kea, Cerro Paranal, and McDonald Observatory, indicating
its broader applicability and robustness in real-world settings.

At the NOT, \vp has become part of the standard operational workflow and is used
routinely for essentially all observations. Its integration with local systems
(e.g. observing block queues) and its role in both preparation and real-time
observations have made it a core component of nightly operations.
Its continuous use over a decade, across evolving
operational conditions and user cohorts, provides strong evidence of reliability
and practical utility in a production environment.
During this period, \vp has been the primary tool for observability assessment
and short-term scheduling, and has therefore been part of the operational
workflow underlying a large body of scientific results\footnote{\url{https://www.not.iac.es/news/publications}}.

\subsection{Educational use}\label{workflow:educational}
\vp is particularly suitable for educational use cases, where instructors or
students can interactively explore target observability and scheduling scenarios,
in order to enhance understanding of astronomical coordinates, airmass,
altitude limits, and the planning process in a classroom or training environment.
Since the software is publicly available, both students and instructors can access it
without requiring a subscription or local installation, which lowers the barrier
to its use in classroom settings. As such, a teacher can have an entire
classroom of students planning an observing night within a few minutes without
debugging Python environments.
The tool also enables visual comparison of target observability across different
times of the year, observatory locations, and facility-specific constraints,
providing a practical framework for introducing the fundamentals of
observational planning.
In addition, \vp helps students understand different representations of a target
on sky by allowing them to compare the altitude plots with the
instantaneous positions in the all-sky display. This combination of graphical
views supports intuitive exploration of concepts such as sky motion, coordinate
systems, and altitude--airmass relationships.

\subsection{Survey results}\label{section:survey}
To assess how \vp is used in practice and to identify its perceived
operational impact, limitations, and areas for future development,
a structured questionnaire (reproduced in \appref{app:surveyquestions}) 
was distributed to known users of \vp.
The survey combined multiple-choice and open-ended questions covering usage context, operational impact, comparison with 
alternative tools, adoption, and teaching applications.
A total of 24 responses were
received from telescope operators, support astronomers, and
visiting observers across multiple facilities.
Because participation was voluntary and targeted toward 
known users, the survey primarily reflects the experiences of professional and
semi-professional astronomers, and should not be interpreted as a representative
sample of the broader astronomical community.

Survey participants reported experience across a wide range of professional astronomical observatories,
with 38\% indicating use at multiple telescopes. 
Most respondents (79\%) 
had used \vp for more than one year, and 79\% reported using it on every observing night or several times per 
week during observing campaigns. The most common applications were nightly scheduling (96\%), 
target visibility assessment (92\%), 
proposal preparation (63\%), contingency planning (54\%) 
, and teaching or training activities (42\%). 

Participants generally reported a high degree of confidence in the software: 88\% indicated that they
use it as the primary 
planning tool, and 92\% reported that
other users at their institution also make use of it.
Adoption appears to have occurred primarily through existing observatory deployments (42\%) and colleague recommendations (38\%). 
Participants frequently reported reductions in planning effort compared
to previous workflows based on spreadsheets, manual calculations, custom scripts, or multiple independent tools.
Reported benefits included faster schedule generation, easier assessment of
visibility constraints, and the ability to evaluate scheduling alternatives interactively.

Survey responses also indicate a mixed deployment model. While 58\% 
of respondents primarily use the public web interface, 38\% 
reported using customized installations integrated with observatory infrastructure,
and 4\% use both. 
This flexibility allows common workflows to be maintained across different facilities while accommodating site-specific
operational requirements.

Teaching and training emerged as a secondary but significant application.
Among the 54\% of respondents who had used \vp in educational settings,
common contexts included observing schools (100\%),
observatory training (77\%),
and undergraduate or graduate instruction (42\%). 
Participants reported that the software was particularly useful for illustrating
core observational concepts such as airmass and visibility constraints (100\%), 
the time-dependent nature of target observability (79\%),
and observation planning trade-offs (71\%). 
All respondents who answered these questions indicated that new users could
learn the interface with minimal (60\%) or moderate (40\%) guidance, 
and that it helps develop intuition for observation planning.

Survey feedback has also contributed directly to the evolution of the software.
Respondents identified desired features, reported operational limitations, and
suggested improvements that have informed subsequent development. Examples
include support for additional telescopes, parallactic-angle calculations, and
external target-name resolution. This feedback-driven development model has
helped ensure that new functionality remains aligned with operational requirements.

\section{Summary and future work}\label{sec:summary}
\vp was developed to fill a specific operational need: a lightweight,
platform-independent tool that bridges the gap between static visibility charts
and complex, high-level robotic schedulers. Since its inception at the NOT
in 2016, the software has evolved into a robust decision-support system.

The distinguishing aspect of \vp is the combination of advanced scheduling
functionality with immediate accessibility.
The software supports flexible, composable, and target-specific observing constraints,
allowing users to define complex observability conditions through combinations 
of airmass and moon distance limits, UTC, LST and HA ranges, etc. These
constraints are evaluated in the context of telescope- and mount-specific limitations,
such as declination-dependent minimum altitudes for equatorial systems or
zenith blind spots for altazimuth mounts, ensuring that generated schedules
are physically executable. The scheduling engine applies a weighted, multi-criteria
heuristic while allowing explicit control over priorities, including the enforced
placement of time-critical or monitoring observations.
This functionality, coupled with a curated database of over 120 
telescopes, automated target resolution via SIMBAD, and automated
non-sidereal ephemeris retrieval via JPL Horizons, allows \vp to function as a 
consistent, portable `mission control' for astronomers working across 
geographically distributed facilities.

While the current version is optimized for independent nightly planning, 
some avenues for future development remain:

\begin{itemize} 
\item Multi-night runs. A natural extension of the current architecture would be
scheduling across multiple consecutive nights, treating an entire observing run
as a unified operational window rather than a series of isolated events.
Targets that remain unobserved due to weather, technical interruptions, or time
constraints would be automatically carried over to subsequent nights, and by
dynamically re-adjusting the priority weights of these missed targets in 
the next night's pool, \vp could ensure the systematic completion of a
scientific program over the course of a run.

\item Automated sky transparency measurements. While \vp currently supports the
visual overlay of targets over all-sky camera images (\secref{workflow:realtime}), future
versions could incorporate automated, real-time estimates of cloudy regions and
sky transparency based on extinction measurements derived from all-sky camera
images (e.g., \citealp{tonry2018}).
Implementation would require observatory-specific calibration because of differences in camera hardware, optics, filters, 
and data-processing pipelines.
\end{itemize}

Ultimately, the guiding philosophy of \vp is to remain lightweight and
easy to use. As \vp continues to refine its utility as a practical bridge
between the planning and execution phases of astronomical observations,
its development will focus on improving reliability and consolidating core
functionality, rather than expanding the tool into a monolithic observatory
control framework.
As an open-source tool, \vp's evolution will continue to be driven by
community feedback.

\begin{acknowledgements}
We thank Anlaug Amanda Djuvpik, John Telting, and the anonymous referee
for their careful reading of the manuscript and valuable feedback.

We thank the members of the \vp user community who completed the survey
and provided detailed feedback on their operational and educational use of the software.
We are particularly grateful to Jos\'{e} F. Ag\"{u}\'{i} Fern\'{a}ndez,
Abel de Burgos, Anlaug Amanda Djupvik, Marcelo A. Fetzner Keniger,
Francisco J. Galindo-Guil, Zuri Gray, Zoe Hackshaw, Roar Holmberg, Ann M. Isaacs,
Carlos Jurado, Anni Kasikov, Emil Knudstrup, Niilo Koivisto, Catherine Manea,
Martin Bo Nielsen, Viktoria Pinter, Tapio Pursimo, Judith Santos-Torres,
Lauri Siltala, Jacco H. Terwel, Mikael Turkki, Kostas Valeckas, and Olga Zamora
for granting permission to acknowledge their participation in the survey by name.

We also thank Peter Sørensen, John Telting, Anlaug Amanda Djupvik, Tapio Pursimo,
Thomas Augusteijn, Ren\'{e} Tronsgaard, Thomas Reynolds, Teet Kuutma,
Andreas Kvammen, and Joonas Saario for their support and feedback during the
initial stages of development and deployment of \vp at the NOT.

Based on observations made with the Nordic Optical Telescope, owned in collaboration 
by the University of Turku and Aarhus University, and operated jointly by Aarhus University, 
the University of Turku and the University of Oslo, representing Denmark, Finland and Norway, 
the University of Iceland and Stockholm University at the Observatorio del Roque de los Muchachos, 
La Palma, Spain, of the Instituto de Astrof\'isica de Canarias.

Based on observations made with the Harlan J. Smith Telescope, owned by the University 
of Texas at Austin at the McDonald Observatory, Texas.

We acknowledge the use of ChatGPT (OpenAI, versions 4--5) for language editing
and clarity suggestions in the drafting of this manuscript.
\end{acknowledgements}

\bibliographystyle{aa} 
\bibliography{aa60375-26}

\begin{appendix}

\section{Example of a full-night schedule}
\begin{center}
\begin{turn}{90}
\begin{minipage}{0.95\textheight}
\centering
\includegraphics[width=.95\textheight]{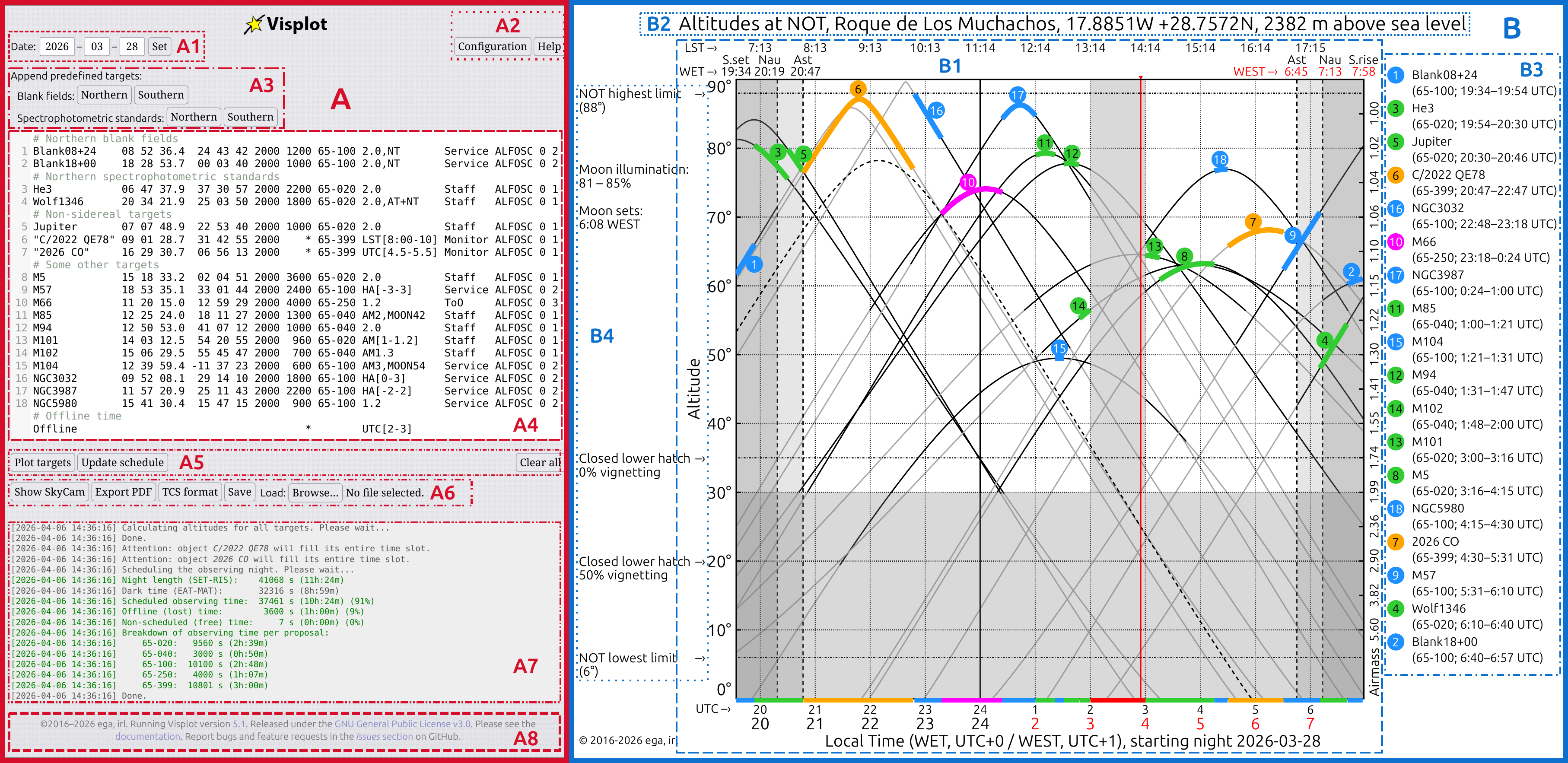}
\captionof{figure}{Example of a full-night schedule produced by \vp for a heterogeneous set of targets.
The interface is organized into two main areas: a configuration panel on
the left (A, red) and a main canvas on the right (B, blue), as discussed in
\secref{workflow:planning}.
The core output of the software are the altitude curves as a function of time,
with dark segments indicating intervals where all constraints are satisfied and light
grey segments where at least one constraint is violated. Different targets are
subject to distinct combinations of constraints (\tabref{table:constraints}),
resulting in different valid observing intervals.
The schedule shown in this figure was produced with the flexible-placement
beam search algorithm (\secref{model:scheduling}), and is discussed in
\secref{workflow:example}.
It includes both sidereal and non-sidereal targets, fixed time constraints,
and a scheduled offline period. The time axis also reflects a transition from WET
to WEST on the given date (marked in red), while UTC and LST remain continuous.
\newline
Note: The target list and proposal identifiers shown here are illustrative and do not correspond to an actual observing night.}
\label{fig:overview}
\end{minipage}
\end{turn}
\end{center}

\clearpage

\section{\vp survey questions}\label{app:surveyquestions}
The survey discussed in \secref{section:survey} was distributed though a Google
form\footnote{\url{https://forms.gle/z6YdLicL9E5U2EEr5}} and contained the
following questions:\\

\noindent A. Context of use

A1. \textit{Please list the observatory(ies) and telescope(s) where you have used Visplot (current or past)}.

A2. \textit{If applicable, which telescope do you currently use Visplot more frequently?}

A3. \textit{How long have you been using/have used Visplot?} (< 6 months / 6-12 months / 1-3 years / > 3 years)

A4. \textit{How often do you use/have used Visplot?} (Every observing night / Several times per week / Occasionally / Other)

A5. \textit{How do you primarily use Visplot?} (Via the public website / I use a locally installed or customized version / Both)

A6. \textit{If you use a local or customized installation, please briefly describe it.}
(For example: local deployment, integration with observing systems, custom constraints, or additional features.)

A7. \textit{In which contexts have you used Visplot?} 
(Nightly observation scheduling / Target visibility checks / Proposal preparation / Backup or contingency planning / Teaching or training / Other)\\

\noindent B. Scientific and operational impact

B1. \textit{Has Visplot changed how you plan observations? If so, how?}

B2. \textit{Compared to your previous approach, Visplot has helped with:}
(Reducing planning time / Improving observing efficiency / Reducing scheduling or visibility errors / Making planning more transparent within the team / Other)

B3. \textit{If possible, can you estimate the impact?}
(e.g. time saved per night, fewer replans, improved target coverage, effort saved, etc.)\\

\noindent C. Comparison with alternatives

C1. \textit{What tools did you use before Visplot?}
(Custom scripts / Spreadsheets / Manual planning / Staralt tool / Other)

C2. \textit{In your experience, what does Visplot do better than those tools?}

C3. \textit{Are there aspects where previous tools performed better?}\\

\noindent D. Trust, adoption, and sustainability

D1. \textit{How did you first learn about Visplot?}
(Recommended by a colleague / Already in use at my telescope or observatory / Found it through a web search / Learned about it via a conference, workshop, or talk / Learned about it via a publication or technical note / Other)

D2. \textit{Do you trust Visplot for operational observing decisions?}
(Yes, as a primary planning tool / Yes, with cross-checks / No, exploratory use only)

D3. \textit{Is Visplot used / has it been used by others at your institution?}
(No, just me / A few people use it / Everybody uses it)

D4. \textit{Have you recommended Visplot to colleagues?}
(Yes / No / Not yet)\\

\noindent E. Teaching and training use

E1. \textit{Have you used Visplot for teaching or training activities?}
(Yes / No)

E2. \textit{If yes, in which context(s) was Visplot used?}
(Undergraduate teaching / Graduate-level teaching / Observing schools or summer schools / Telescope or observatory training / Outreach or informal education / Other)

E3. \textit{How effective was Visplot for teaching or training?}
(1 Not very effective ... 4 Very effective)

E4. \textit{In your experience, did Visplot help learners grasp any of the following concepts?}
(Astronomical coordinate systems / Altitude--azimuth and sky motion / Visibility constraints (airmass, horizon, twilight) / Time-dependent observability / Observation planning trade-offs / None in particular / Other)

E5. \textit{Was Visplot easy for students or trainees to understand and use?}
(Yes, with minimal guidance / Yes, with some guidance / No, it required significant guidance)

E6. \textit{Would you consider using Visplot again for teaching or training?}
(Yes / No)

E7. Optional comments on teaching or training use.
\textit{Please share any comments on how Visplot was used for teaching or training, including what worked well or what could be improved.}\\

\noindent Final Section. Additional comments.

\textit{Please use this space to share any additional comments that were not covered above.
This may include suggestions for improvement, limitations you have encountered, features you find particularly valuable, or any other feedback you think is relevant for documenting Visplot's scientific and operational use.}
\end{appendix}

\end{document}